\documentclass{iaus}
\usepackage{graphicx}

\title[Star Formation in Disks] 
{Star Formation in Disks: Spiral Arms, Turbulence, and Triggering
Mechanisms}

\author[Bruce G. Elmegreen]   
{Bruce G. Elmegreen$^1$ \affiliation{$^1$IBM Research Division, T.J.
Watson Research Center, 1101 Kitchawan Road, Yorktown Hts. 10598, USA
\\ email: {\tt bge@us.ibm.com}}}

\pubyear{2009}
\volume{254}  
\pagerange{1-11}
 \jname{The Galaxy Disk in Cosmological Context}
\editors{J. Andersen, J. Bland-Hawthorn \& B. Nordström, eds.}
\begin{document}

\maketitle

\begin{abstract}
Star formation is enhanced in spiral arms because of a combination of
orbit crowding, cloud collisions, and gravitational instabilities. The
characteristic mass for the instability is $10^7\;M_\odot$ in gas and
$10^5\;M_\odot$ in stars, and the morphology is the familiar beads on a
string with 1-2 kpc separation. Similar instabilities occur in
resonance rings and tidal tails. Sequential triggering from stellar
pressure occurs in two ways. For short times and near distances, it
occurs in the bright rims and dense knots that lag behind during cloud
dispersal. For long times, it occurs in swept-up shells and along the
periphery of cleared regions. The first case should be common but
difficult to disentangle from independent star formation in the same
cloud. The second case has a causality condition and a collapse
condition and is often easy to recognize. Turbulent triggering produces
a hierarchy of dense cloudy structure and an associated hierarchy of
young star positions. There should also be a correlation between the
duration of star formation and the size of the region that is analogous
to the size-linewidth relation in the gas. The cosmological context is
provided by observations of star formation in high redshift galaxies.
Sequential and turbulent triggering is not yet observable, but
gravitational instabilities are, and they show a scale up from local
instabilities by a factor of $\sim3$ in size and $\sim100$ in mass.
This is most easily explained as the result of an increase in the ISM
turbulent speed by a factor of $\sim5$. In the clumpiest galaxies at
high redshift, the clumps are so large that they should interact with
each other and merge in the center, where they form or contribute to
the bulge. \keywords{stars: formation, ISM: evolution, galaxies:
formation, galaxies: high-redshift}
\end{abstract}

\firstsection 
\section{Star Formation in Spiral Arms}

Star formation in local disk galaxies, including the Milky Way, is
highly concentrated in spiral arms. This is partly because the gas is
concentrated in the arms as a result of density-wave streaming. The
stellar spiral arm potential pulls out the interarm gas nearly
radially, causing it to traverse the interarm region quickly, and it
shocks and pulls back the gas when it enters the arm, causing it to
deflect and move nearly parallel to the arm. Because the time spent at
each position relative to an arm is inversely proportional to the
perpendicular component of the velocity, and because the local density
is also inversely proportional to this velocity, the gas lingers in an
arm for a time proportional to its density.  Thus most star formation
occurs where the arm is dense because most of the time and most of the
gas occurs there. If this were all that happened, then the star
formation rate should be directly proportional to the total gas density
in an azimuthal profile around the disk. There are no recent studies of
the azimuthal dependence of star formation, although more on this topic
was done over a decade ago (e.g., Garcia-Burillo et al. 1993). In
addition to this purely kinematical effect, there may also be a
dynamical effect in the sense that the star formation rate per unit gas
mass increases in the arms relative to the interarm regions. In this
case, the star formation rate should increase with the total gas
density to a power greater than unity for azimuthal profiles. In either
case, the sizes of the star-forming regions are much larger in the arms
than between the arms (e.g., Lundgren et al. 2004).

Consideration of dynamical processes and the time available suggest
that spiral arms could promote local gravitational instabilities that
trigger star formation in giant complexes (e.g., Tosaki et al. 2002).
In the low-shear environment of an arm (e.g., Luna et al. 2006), such
instabilities produce $10^7\;M_\odot$ clumps rather than spiral
wavelets. Magnetic fields help by removing gas angular momentum during
subsequent collapse (Kim \& Ostriker 2002). The morphology of star
formation then consists of HI clouds with molecular cores lining the
stellar spiral arms (e.g., Engargiola et al. 2004). Feathers and spurs
extend into the interarm region as the arm clouds feel an increased
shear (e.g., La Vigne et al. 2006). Deep in the interarm regions, star
formation lingers in the long-lived envelopes and diffuse debris of
spiral arm clouds (Elmegreen 2007).

In galaxies without strong stellar spirals, star formation occurs
throughout the disk in flocculent arms that are probably the result of
local, swing-amplified instabilities (e.g., Fuchs et al. 2005). Thus,
star-forming clouds form or grow in (1) spiral-wave triggered
gravitational instabilities and cloud collisions in dense dust lanes,
and (2) random gravitational instabilities everywhere if there are no
stellar spirals, or if the stellar spirals are weak. In either case,
star formation follows cloud formation in a few cloud dynamical times.
It lingers and gets triggered at a low level in the molecular cloud
debris for a much longer time.

\section{Triggering}
Triggering of star formation is well known in places like the pillars
of M16 (Hester et al. 1996), and other bright rims (e.g., Sugitani et
al. 1989; Reach et al. 2004). The large-scale structure of Ophiuchus
also suggests that the main star formation was triggered in the head of
a cometary cloud shaped by pressures from the Sco-Cen OB association
(de Geus 1992). Other triggering takes the form of shells, with old
stars in the center and young stars along the periphery. Such shells
have been observed in local star-forming regions (e.g., Zavagno et al.
2006) and in other galaxies (e.g., Brinks \& Bajaja 1986). Some
triggering shells can be very large (Wilcots \& Miller 1998; Walter \&
Brinks 1999; Yamaguchi et al. 2001a) and some contain triggered pillars
also (e.g., Yamaguchi et al. 2001b; Oey et al. 2005).

Generally, the spatial scale for triggering is the shock speed in the
ambient gas multiplied by the timescale for the pressure source,
$L\sim(P/\rho_0)^{1/2}\times T$. If the spatial scale is smaller than
the cloud in which the first generation of star formation occurs, then
pillars and bright rims form by the push-back of interclump gas. Star
formation triggering can be fast in this case, prompted by the direct
squeezing of pre-existing dense gas. Also in this case, the relative
velocity of the triggered stars will be small because they are in the
dense gas that is left behind. If the triggering scale is larger than
the size of the cloud, then shells form by the push-back of all the
gas, both in the cloud and in the intercloud medium. This is a slow
process because new clumps have to form, and the time scale for this is
$\sim(G\rho_{shell})^{-1/2}$. The velocity of triggered stars in this
second case can be large, comparable to the shock speed,
$\sim(P/\rho_0)^{1/2}$. There is also a causality condition, where the
triggering distance between generations equals the time difference
multiplied by the velocity difference.  Criteria for gravitational
collapse in expanding shells were derived by Elmegreen (1994),
Elmegreen, Palous \& Ehlerova (2002), and Whitworth et al. (1994).

There are many types and sources of energy in the ISM and many possible
causes for triggering. Energy {\it types} include thermal, magnetic,
turbulent, cosmic ray, and rotational. The first four are all
comparable and equal to several tenths of an eV cm$^{-3}$. Rotational
energy is much denser, several hundred eV cm$^{-3}$. Of the first four,
only supersonic turbulence has been associated with star formation
triggering because such turbulence can compress the gas a lot. Thermal
instabilities compress the gas too, but usually in small regions where
self-gravity is not important. Magnetic fields cause slight compression
when the gas rearranges itself on the field lines following a Parker
instability, but this rearrangement alone is not enough to trigger star
formation -- that requires self-gravity too. Cosmic rays interact with
the ISM primarily through heating and the generation of small scale MHD
turbulence. If this turbulence compresses the gas significantly, and on
sufficiently large scales, then cosmic rays could trigger star
formation (there has been no theoretical work on this mechanism as far
as I know). Rotational energy compresses the gas and triggers star
formation more than any of the others when something gets in the way of
the uniform circular motion. Then a galactic-scale shock forms and star
formation results in the dense gas. Spiral density waves in the stars
and stellar bars can do this, as mentioned in the previous section.

Among the various {\it sources} of ISM energy, we include supernovae,
HII regions and stellar winds as stellar sources, and galaxy rotation
as an energy source for the magnetic field through the dynamo. The
magnetic field drives turbulence and convection into the halo by the
Parker instability (e.g., Kosi\'nski, \& Hanasz 2006; Lee \& Hong
2007), and it drives turbulence in the midplane by amplifying epicyclic
oscillations in the radial direction, which is the Magneto-Rotational
instability (e.g., Kim, Ostriker, \& Stone 2003). Galactic rotation
also generates turbulence at spiral waves (Bonnell, et al. 2006; Kim,
Kim, \& Ostriker 2006; Dobbs \& Bonnell 2007). In addition, there is
star-light energy, which is comparable in density to turbulence, cosmic
rays, and other energies mentioned above, but is not so easily coupled
to the gas. There is also ISM self-gravity, which may be considered as
a source of energy for motions if it is replenished by cloud
disruption, in which case the real energy source is that of the
disruption (i.e., young stars; self-gravity only stores the other
energies in potential form.)

Of these sources, most do not trigger star formation because they do
not interact with gas in the right way. The most important energy
sources are those that produce high pressures for long times -- long
enough for gravity to act in the compressed regions. Single supernovae,
for example, are short-lived, with radiative lifetimes only a percent
of the local dynamical time of the gas they are in (Dekel \& Silk
1986). Magnetic energy, cosmic rays, and starlight do not compress the
gas much. HII regions and O-star winds compress only the most local
gas, but they do this for a long time and often trigger star formation
in adjacent molecular clouds. Combined pressures from HII regions,
winds and supernovae can cause major triggering: they act in OB
associations for a relatively long time, $\sim5$ Myr or more, and can
trigger star formation on a scale of 10 to 100 pc or more in the
surrounding gas, which includes remnants of the molecular clouds that
formed these stars, neighboring molecular clouds, and intercloud gas.

In summary, star formation either follows cloud formation, or it is
stimulated in existing clouds by external processes.  Gaseous
self-gravity alone triggers cloud formation through: (1) dust-lane
fragmentation in stellar density waves, (2) ring fragmentation in
Lindblad resonance rings, (3) tidal clump or dwarf galaxy formation in
the tidal arms of interacting galaxies, and (4) swing amplified clumps
if there are no imposed stellar structures. Self-gravity also causes
existing clouds to collapse and fragment into denser pieces where stars
form. The morphologies of these processes are relatively easy to
recognize in ideal cases: they appear as ``beads on a string'' of star
formation in spiral arms, resonance rings, and tidal tails, or they are
sheared spiral-like clumps from local instabilities in the gas. All of
the clouds or cores formed by self-gravity have masses exceeding the
relevant Jeans mass. In the case of instabilities in disks, filaments
and shocked layers, the elongated regions are always unstable to
condensation into spheroids. If the resulting spheroids are massive
enough, then they can collapse further into clusters or single stars.
If the disk is rotating, or the layer is expanding, then the largest
scales are stabilized and there is a column density threshold that has
to be exceeded so that regions large enough to exceed the Jeans length
(where gravity exceeds pressure) are also smaller than the
stabilization length (where rotation or expansion exceed gravity).

Inside the clouds formed by these processes, star formation is also
triggered by locally high pressures from HII regions, winds, and
multiple SNe, and from cloud collisions, or in supersonically turbulent
media, by compressions from converging flows. All of these compressions
tend to enhance magnetic diffusion and decrease the dynamical time. The
morphology of this triggering is also fairly easy to recognize in ideal
cases because the compressed regions are shells, comets, and moving
layers, all adjacent to high pressures, and all with the causality
constraint mentioned above.

\section{Turbulence Triggering}

The compressed regions in a supersonically turbulent cloud act like
seeds for gravitational attraction and can lead to the local accretions
necessary to make stars (see reviews in Mac Low \& Klessen 2004;
Bonnell et al. 2007). Simulations by several groups have shown how
turbulence in a cloud core can produce a star cluster (e.g., Li et al.
2004; Bate \& Bonnell 2005; Jappsen et al. 2005; Padoan et al. 2005;
Nakamura \& Li 2005; Martel et al. 2006; Tilley \& Pudritz 2007).

There are several signatures of turbulence triggering. First, the
cloudy structure in a turbulent medium has a power law power spectrum,
which means there is no characteristic scale except the largest scale
(e.g., St\"utzke et al. 1998). In fact, the morphology tends to be
hierarchical, with large clouds containing small subclouds over many
levels. This hierarchy in gas produces a similar hierarchy in young
stars, which is evident as substructure in embedded clusters (e.g.,
Testi et al. 2000; Dahm \& Simon 2005; Gutermuth et al. 2005), and as
nested super-structures on larger scales (e.g., several subgroups are
collected into each OB association, and several OB associations are
collected into each star complex; see review in Elmegreen 2008a). Each
galactic cluster seems to be the inner mixed region of the hierarchy,
where the efficiency of star formation is automatically high (Elmegreen
2008b).

Second, the hierarchy of structures produces a mass spectrum for
clusters, and nearly the same mass spectrum for stars, that is
$dN/dM\propto M^{-2}$ (the Salpeter IMF would have a power $-2.35$).
This is because each layer in the hierarchy contains the same total
mass, just divided up in different ways. Because the hierarchy is a
sequence in the $\log$ of the mass (each level divides up the mass of
the previous higher level), we have the mass conserving requirement
that $MdN/d\log M =$constant, that is, the total mass in each $\log M$
interval is constant. This converts to $M^2dN/dM=$constant, as above.
Another way to view this is to consider a hierarchy of levels where the
final product of star formation, a cluster, for example, or an OB
association, comes from some cloud at one of many possible levels in
the hierarchy. That cloud is contained, along with other clouds, in the
next higher cloud, and also is subdivided into several sub-clouds. Then
it turns out that the same mass function follows by randomly sampling
all clouds in the hierarchical tree. That is, if every cloud (and all
of its subclouds) is equally likely to produce a stellar object as
every other cloud, then $M^2dN/dM=$constant again. This may be seen
with a simple example. Imagine clouds subdivided by twos: one cloud of
mass 32 units divided into 2 clouds of mass 16 units, which are each
divided into 2 more clouds of 8 units, and so on until the smallest
level, which has 32 clouds of 1 unit mass each. The total number of
clouds, counting everything, is $32+16+8+4+2+1=63$. The probability
that a cloud of mass 8, say, is selected, is the number of clouds with
mass 8, namely 4 clouds, divided by the total number of clouds, 63. The
probability that a cloud of mass 4 is chosen is similarly 8/63. In
general, the probability that a cloud of mass $M$ is chosen is
$\propto1/M$. Since we have intervals of $\log M$ again, and we are
assuming the number of clouds is proportional to their probability, we
get $dN/d\log M\propto 1/M$, or $dN/dM\propto 1/M^2$, as above.  Thus
turbulence, and the hierarchical structure it always produces, is the
likely cause of the $M^{-2}$ mass functions for clusters and OB
associations (Elmegreen \& Efremov 1997), and maybe even stars.

The mass function for clouds is measured differently. Clouds are
usually defined by the resolution of a survey. That is, most cloud mass
functions have clouds with sizes within a factor of 10 of the survey
resolution. The total span of mass is therefore only a factor of 100
(since mass is proportional to size-squared). Larger clouds are not
counted because they can always be subdivided into their smaller
pieces. This is a problem with defining clouds as regions inside of
closed contours, for example, and of disallowing any multiple-counting
of mass. Turbulent media are not really composed of separate clouds
within an intercloud medium. Contours do not represent the power law
structure correctly and mass spectra obtained from contours do not
represent the true distribution of mass into all of its parts. For
stellar structures, we can measure objects arbitrarily large (unlike
the case for contoured clouds) because the stellar objects are not
defined by their structure (e.g., contours) but by their stellar
content: IR-excess stars, for example, define a young cluster, OB stars
define an OB association, and red supergiants define a star complex
(Efremov 1995). This connection between stellar types and structure
nomenclature is a selection effect, resulting from the third aspect of
turbulence triggering, discussed next. Also, stellar aggregates tend to
be defined by friends-of-friends algorithms, which is intrinsically
hierarchical, unlike contouring. Better algorithms for counting clouds
have recently been devised (Rosolowsky et al. 2008).

Third, turbulence triggering tends to occur on the timescale for the
turbulent motions to move through a region, i.e., the crossing time.
Because the turbulent speed scales with a fractional power of the size,
the crossing time (size divided by speed) also increases with a
fractional power of the size. This means that larger regions form stars
longer. This correlation has been observed in the LMC clusters and
cepheids (Efremov \& Elmegreen 1998).  As a result, regions defined by
stars with a certain age range, such as IR-excess stars, O-type stars,
etc., tend to have a certain size and mass. They are not distinct
objects, however. OB associations are not intrinsically different from
T associations or embedded clusters or star complexes, aside from their
difference in selected age range. Recall that turbulence is scale free.
It is only the selection of an age that corresponds to the selection of
a certain scale or mass of star formation. This is true up to the
largest scale for star formation, which is a flocculent spiral arm or
one of the beads-on-a-string in a stellar arm (Elmegreen \& Efremov
1996; Odekon 2008).

Turbulence triggering seems to work along side sequential triggering on
scales smaller than the ambient Jeans length, which is $L_J$ defined
below. This is also about the galactic scale height. Thus a simplified
model for all of these processes would be that gravitational
instabilities produce clouds and drive turbulence on the largest scales
($L_J$), while turbulence and gravitational collapse trigger the first
generation of stars inside these primary clouds. Sequential triggering
prolongs star formation in the debris during the process of cloud
disruption. The fractions of stars triggered by turbulence and by the
various sequential processes would be interesting to observe.

\section{The Cosmological Connection}

Deep surveys in the optical and infrared have produced images of
thousands of young galaxies, some of which are half, or even one-tenth,
the age of the current universe. It is interesting to ask whether star
formation in these galaxies has the same cause and morphology as local
star formation.

To investigate this, we have measured regions of star formation in all
of the large ($>10$ pixels) galaxies in the Hubble Space Telescope
Ultra Deep Field (UDF, Beckwith 2006). There are about 1000 of them
(Elmegreen et al. 2005). Galaxies look different at high redshift.
Irregular structures dominate, and interactions are relatively common
(Abraham et al. 1996a,b; Conselice, Blackburne, \& Papovich 2005). The
one irregularity that seems to be most common is the presence of giant
blue clumps from star formation (Elmegreen \& Elmegreen 2005). These
clumps are usually observed in the restframe ultraviolet because of the
high redshifts involved, but even there they have absolute magnitudes
that can be brighter than $-18$ mag (Elmegreen \& Elmegreen 2006b;
Elmegreen et al. 2007a, 2008c). Population models suggest that clump
stellar masses are in the range $10^7-10^8\;M_\odot$, with some clumps
larger than $10^9\;M_\odot$ (Elmegreen et al. 2007a, 2008c). Clump
diameters are $\sim1.5$ kpc, and stellar ages in the clumps are
typically within a factor of 3 of $10^8$ yrs (Elmegreen \& Elmegreen
2005; Elmegreen et al. 2008c).

There are no obvious shells or comets of triggered star formation at
high redshift, but none are expected at the available resolution of
$\sim200$ pc out to $z\sim5$. In the most clumpy galaxies, which are
the {\it clump clusters} and {\it chains} (Elmegreen et al. 2005),
there are no spirals either, even though spiral structure would be seen
at $\sim5$ if it was bright enough in the uv. Other galaxies clearly
have spirals and bulges. At $z>1$, clump clusters and chains in the UDF
outnumber spirals 2:1.

In a recent survey of bulge properties using H-band NICMOS observations
in the UDF (Elmegreen et al. 2008c), we found that $\sim50$\% of the
clump clusters and $\sim30$\% of the chains have bright, often central,
red clumps indicative of bulges. Others may have only a red and smooth
underlying disk. We also found that when there are bulges in the most
clumpy galaxies, these bulges are more similar to the clumps with
respect to age and mass than the bulges in spiral galaxies are to their
clumps. Thus clump clusters and chains either have no bulges or they
have young bulges.

\begin{figure}[b]
\begin{center}
 \includegraphics[width=5.in]{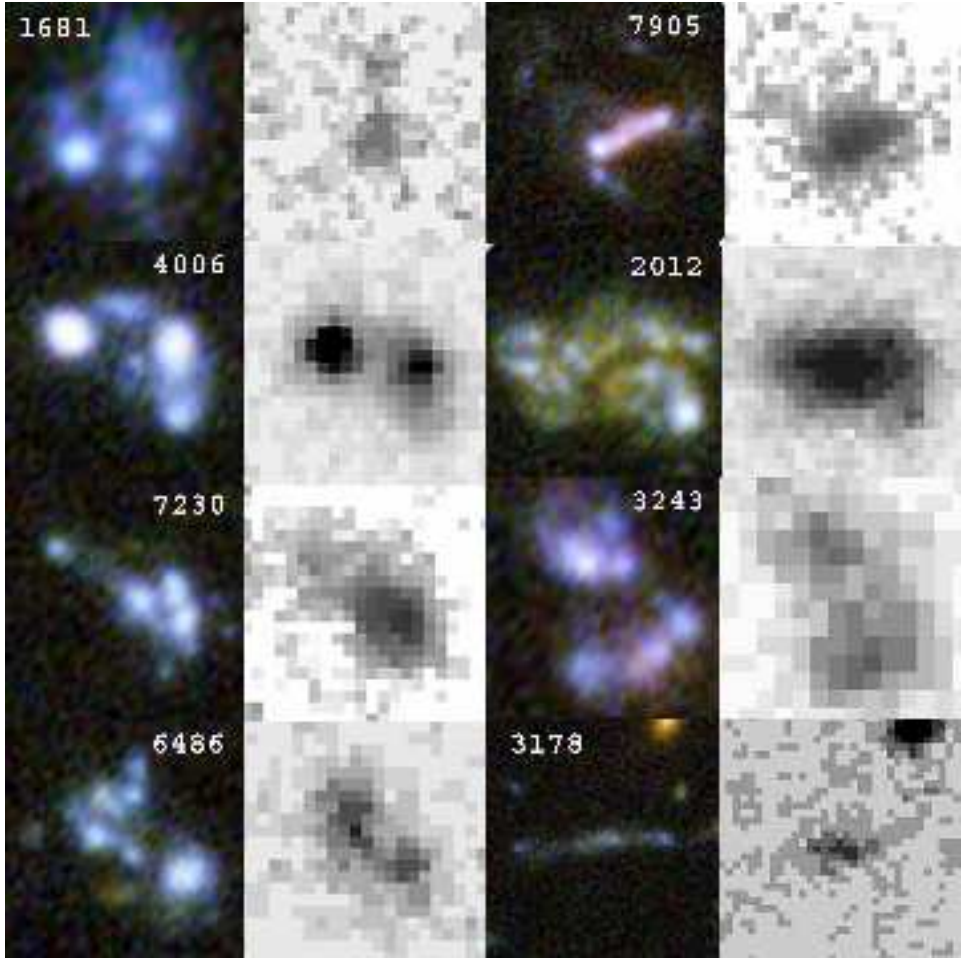}
\caption{Eight clumpy galaxies in the UDF with no obvious central red
object that might be a bulge. Each galaxy is shown twice, on the left
using the high resolution of the ACS from the color Skywalker image,
and on the right using the lower resolution NICMOS image in H band. Red
emission is either extended in a seemingly old population of stars, or
it is associated with bright star-forming regions. (Image degraded for
astro-ph)}
\end{center}
\end{figure}

Figure 1 shows a selection of UDF clump cluster galaxies without
obvious bulges (from Elmegreen et al. 2008c, which contains a color
version of this figure). Each galaxy is shown as a pair of images, with
the color Skywalker image on the left and the NICMOS H-band image on
the right. There are bright clumps at NICMOS-H, but nothing centralized
and nothing obviously connected with a disk population; most are
associated with a star formation feature, and some could be independent
galaxies. Either these clump clusters will turn into late-type spirals
with small bulges, or they have not yet formed their bulges.
Conceivably, one of the prominent clumps could migrate to the center to
make a bulge, or several could collide and make a bulge. This
clump-migration model for bulge formation was proposed by Noguchi
(1999) and Immeli et al. (2004a,b). More details and a match to
observations is in recent papers by Bournaud et al. (2007, 2008) and
Elmegreen et al. (2008a,c).

The recent batch of models starts with a disk that is half gas and half
stars, in addition to a live halo. The stellar part is Toomre-stable
but the gas+stellar disk is unstable. Almost immediately, the gas and
some of the stars collapse into four or five giant clumps with
$10^8\;M_\odot$, like the observed clumps in clump-cluster galaxies.
When the disk is slightly off-center from the bulge, the clumps get
flung around to large radii, as observed in UDF 6462 (Bournaud et al.
2008), contributing to the general appearance of irregularity. The
basic process of clump formation is a gravitational instability. The
collapse is rapid and it forms clumps rather than spirals because the
self-gravitational forces dominate the background galactic forces. This
result seems reasonable for a young galaxy primed with fresh gas from
cosmological accretion. The instability has two important differences
compared with that in local galaxies: first, the high gas fraction
makes the clumps round rather than spiral-like, and second, a high
turbulent speed makes the clumps massive. Recall that the bulk Jeans
length is $L_J=2\sigma^2/G\Sigma$ for turbulent speed $\sigma$ and disk
column density $\Sigma$, and the Jeans mass is
$M_J=\sigma^4/G^2\Sigma$. The key to a large unstable mass is a large
velocity dispersion.

The basic scales $L_J$ and $M_J$ come from the dispersion relation for
gravitational instabilities in an infinitely thin and extended disk,
which is $\omega^2=\sigma^2 k^2 - 2\pi k G \Sigma$; $i\omega$ is the
growth rate and $k$ is the wavenumber. The wavenumber at fastest growth
is obtained by setting $d\omega/dk=0$, which gives $k_{fast}=\pi
G\Sigma/\sigma^2$. The Jeans length is $L_J=2\pi/k_{fast}$ and the
Jeans mass is $M_J\sim\left(L_J/2\right)^2\Sigma=\sigma^4/G^2\Sigma$,
as written above. This mass could be written in a variety of ways, such
as $\pi\left(L_J/2\right)^2\Sigma$ or $\left(L_J\right)^2\Sigma$,
depending on assumptions about what constitutes a cloud; i.e., what
fraction of the unstable mass gets into the cloud. The preferred form
is a compromise, and selected partly to match local observations of
giant cloud masses. Additional things modify the dispersion relation,
such as magnetic fields, spiral arms, finite disk thickness, turbulent
motions, a non-isothermal equation of state, molecule formation, and so
on. Simulations reproduce many of these effects in a way that simple
expressions cannot. A detailed model for star formation in a galaxy
disk, such as that by Robertson \& Kravtsov (2008) discussed at this
conference, should do a better job of defining a characteristic scale
or outer-scale for cloud formation.  Still, it is useful to see how the
basic observed quantities, like mass and length, vary with ISM
properties.

The mass and size of star formation clumps in high redshift galaxies
exceed those in local galaxies by factors of $\sim100$ and $\sim3$,
respectively. Thus $\sigma^2$ has to increase by $100/3\sim30$, which
means that $\sigma$ has to increase by $\times5.5$, making it $30$ or
$40$ km s$^{-1}$ instead of the local $6$ or 7 km s$^{-1}$. This
requirement is satisfied by the observation of high velocity
dispersions in high redshift galaxies (F{\"o}rster Schreiber, et al.
2006; Genzel et al. 2006, 2008; Weiner et al. 2006). Similarly,
$\Sigma$ has to be larger by a factor of $100/3^2\sim10$ at high
redshift. The gaseous disks of local spiral galaxies have
$\Sigma\sim10\;M_\odot$ pc$^{-2}$, which makes
$M_J\sim2\times10^7\;M_\odot$, comparable to the observed local gas
clump mass, for $\sigma=6$ km s$^{-1}$. At high redshift, the unstable
column density in the disk has to be $\sim100\;M_\odot$ pc$^{-2}$,
which is comparable to the total mass column density of the inner
regions of today's spirals.  Considering that high redshift spiral
galaxies are slightly smaller than today's spirals (Elmegreen et al.
2007b), the clump clusters and chains could be forming the inner thick
disks and bulges of today's galaxies.

A big uncertainty for star formation studies of high redshift galaxies
is the neutral gas abundance. CO has been observed in several galaxies
(Solomon \& Vanden Bout 2005; Tacconi et al. 2008) but with little
resolution into clouds. HI has not been observed in emission yet and
the absorption of HI, in the form of damped Lyman $\alpha$ lines, has
an unknown geometry relative to stellar galaxies (Wolfe et al. 2008).
CO is also present in DLA gas (Srianand et al. 2008). Quite possibly,
the gas mass is comparable to or larger than the observed stellar mass,
and the gas is as irregular as the stars in these clumpy disks. Our
prediction is that the velocity dispersion of the neutral gas should be
high, something like $40$ km s$^{-1}$ or more, which is observed for
the ionized gas. The dispersion inside the clumps should be high also,
although perhaps not quite as high if only the cores are observed in
molecular transitions.

There are two other pieces of evidence that star formation is prompted
by gravitational instabilities in high redshift disks. First there is a
linear alignment of clumps in chain galaxies, which are presumably
edge-on clumpy disks. Clump positions are aligned along the midplane of
the chain to within a fraction of a pixel (Elmegreen \& Elmegreen
2006a) or $\sim 100$ pc, on average (i.e., for 112 chain galaxies in
the UDF). This implies that most clumps are not extragalactic objects
in the process of coalescence; they formed in a pre-existing disk.
Second, a large fraction of interacting galaxies with tidal features
and rings have regularly spaced clumps in those features (Elmegreen et
al. 2007a). They therefore had to form there, and their separation
should be comparable to $L_J$. This gives the same requirement on
velocity dispersion as the clumps in non-interacting disks.

\section{Conclusions}

Giant cloud formation is often triggered by gravitational and
associated instabilities in gas disks. The clump scale is $\sim600$ pc
for local galaxies, and what forms is a ``star-complex,'' composed of
OB associations and dense clusters with $10^5-10^6\;M_\odot$ of stars.
The cloud mass at the beginning of the process is $\sim10^7\;M_\odot$.
Most of this gas is in the form of low-density HI except in the dense
inner regions of galaxies, where it can be largely molecular because of
the higher ambient pressure.

In high-redshift disks, the clump scale is larger, $\sim1500$ pc, and
the stellar mass is larger, $\sim10^7-10^8\;M_\odot$. The associated
gas mass in a clump is unknown but may be $10^8-10^{9}\;M_\odot$ with
much of that molecular. This scale-up of star formation at high $z$
seems to be the result of a high turbulent speed, which, combined with
a higher gas column density, makes the gravitational length and
characteristic mass larger by factors of $\sim3$ and $\sim100$,
respectively.

In local galaxies, star formation begins quickly in cloud cores once
they become self-gravitating. There is no reason to think otherwise for
high redshifts. Locally, the cloud cores are cold, dense, molecular,
magnetic, and turbulent -- all necessary attributes contributing to
star formation in one way or another (including the requirement of
angular momentum loss during star formation). The same should be true
at high redshift too.  Outside the local dense cores, star formation
lingers in isolated cloud debris and it can be triggered for a long
time in the dispersing cloud envelopes. The analogous final stages of
star formation at high redshift are unknown. Cloud dispersal by star
formation feedback should be more difficult with a higher velocity
dispersion (Elmegreen et al. 2008b). If this difficulty increases the
efficiency of star formation, then comparatively little gas could
remain in the cloud envelopes for prolonged triggering. Generally,
cloud envelopes are more stable than their cores because of the
envelope exposure to background starlight and the resulting longer
magnetic diffusion time. This is why triggering can be important in the
final stages of cloud disruption: compression enhances the diffusion
rate while it shortens the dynamical time, causing low density regions
to form stars where they otherwise would not. There is no evidence for
sequentially-triggered star formation at high redshift, however, but
the resolution is not good enough yet to see it.

Other differences between low and high redshift star formation concern
the fate of the stellar clumps. At low z, where the disk gas fraction
is low and the clump formation and evolution times are comparable to
the shear time, star-forming clumps dissolve slowly into star streams
and add to the thin disk without changing their galactocentric radii.
At high z, clump formation appears to be much more violent and rapid
compared to background galactic processes. This is observed directly in
clump clusters and chain galaxies, and it is true to a certain degree
also in spirals, where the clumps are more massive than they are
locally. These differences follow from the observed relatively high
turbulent speed and the expected high gas mass fraction. Simulations
suggest that the clumps in the most clumpy disks interact with each
other, stir the preexisting stars to make a thicker disk, shed half of
their own stars during these interactions to add to this thick disk,
and then spiral in the remaining half to make or add to a bulge.

The general theme of this conference is to understand the formation and
structure of the Milky Way in the context of various models and
observations of galaxies on cosmological scales.  In terms of star
formation, the youngest resolvable disk galaxies look both strange, in
terms of their increased clumpiness, and familiar, in terms of the
likely processes involved.  Simulations can reproduce the basic
structures without difficulty if they start with ideal initial
conditions. At some point, however, the galaxy formation process has to
be important, and this involves the gas accretion rate and geometry,
and the galaxy interaction rate. It is more difficult to understand
these aspects of young galaxies without the local analogues that are so
revealing for star formation.

\end{document}